\documentclass[letterpaper, 10 pt, conference]{ieeeconf}  % Comment this line out if you need a4paper

\IEEEoverridecommandlockouts                              % This command is only needed if 
\usepackage[utf8]{inputenc}
\usepackage{dsfont}
\usepackage[usenames,dvipsnames]{xcolor}
\usepackage{graphicx} % for pdf, bitmapped graphics files
\usepackage{mathptmx} % assumes new font selection scheme installed
\usepackage{times} % assumes new font selection scheme installed
\usepackage{amsmath} % assumes amsmath package installed
\usepackage{amssymb}  % assumes amsmath package installed
 
\usepackage{amsthm}
\usepackage{pstricks,pst-plot,psfrag}
\usepackage{multirow}
\usepackage{subcaption}
\usepackage[draft]{hyperref}
\usepackage{subcaption}
\usepackage{graphicx}
\usepackage{environ}         % provides \BODY
\usepackage{etoolbox}        % provides \ifdimcomp

\usepackage[disable]{todonotes}
\usepackage{optidef}
\usepackage[ruled, lined, linesnumbered, shortend]{algorithm2e}
\usepackage[dvipsnames]{xcolor}

\theoremstyle{definition}
\newtheorem{definition}{Definition}[section]
\newtheorem{theorem}{Theorem}[section]
\newtheorem*{theorem*}{Theorem}
\newtheorem{corollary}{Corollary}[theorem]
\newtheorem*{corollary*}{Corollary}
\newtheorem{lemma}[theorem]{Lemma}
\newtheorem*{lemma*}{Lemma}

\newcommand{\setdiff}[0]{\setminus}
\newcommand{\norm}[1]{\left\lVert#1\right\rVert}

\newtheorem{innercustomthm}{Theorem}
\newenvironment{customthm}[1]
  {\renewcommand\theinnercustomthm{#1}\innercustomthm}
  {\endinnercustomthm}

\let\oldnl\nl% Store \nl in \oldnl
\newcommand{\nonl}{\renewcommand{\nl}{\let\nl\oldnl}}% Remove line number for one line

\newif\ifextendedv % Toggle appendix and references to "Supplementary Material" vs no appendix and references to "Extended Version"
\extendedvtrue

\newif\ifmargincomments %A quick way of turning off margin comments for, say, arXiv submission
\margincommentsfalse
\ifmargincomments

\newcommand{\frmargin}[2]{{\color{brown}#1}\marginpar{\color{brown}\raggedright\footnotesize [FR]:#2}}

\newcommand{\ksmargin}[2]{{\color{cyan}#1}\marginpar{\color{cyan}\raggedright\footnotesize [KS]:#2}}

\else

\newcommand{\frmargin}[2]{#1}

\newcommand{\ksmargin}[2]{#1}

\fi

\newif\ifrelaxedv  % Relaxed version with no spacing tricks
\relaxedvtrue
\ifrelaxedv
\newcommand{\jvspace}[1]{}
 \else
\renewcommand{\baselinestretch}{0.96}
\allowdisplaybreaks
\newcommand{\jvspace}[1]{\vspace{#1}}
\fi

\title{\LARGE \bf
On Local Computation for Optimization in Multi-Agent Systems}
\author{Robin Brown$^{1}$, Federico Rossi$^{2}$, Kiril Solovey$^{3}$, Michael T. Wolf$^{2}$, and Marco Pavone$^{3}$% <-this % stops a space
\thanks{$^{1}$R. Brown is with the Institute for Computational \& Mathematical Engineering, Stanford University, Stanford, CA, 94305, {\tt rabrown1@stanford.edu}.}
\thanks{$^{2}$F.\ Rossi, and  M.\ T.\ Wolf are with the Jet Propulsion Laboratory, California Institute of Technology, Pasadena, CA, 91109, {\tt \{federico.rossi, michael.t.wolf\}@jpl.nasa.gov}.}
\thanks{$^{3}$K.\ Solovey and M.\ Pavone are with the Department of Aeronautics \& Astronautics, Stanford University, Stanford, CA, 94305, {\tt \{kirilsol,pavone\}@stanford.edu}.}
}

\begin{document}

\maketitle
\begin{abstract}
A number of prototypical optimization problems in multi-agent systems (e.g., task allocation and network load-sharing) exhibit a highly local structure: that is, each agent's decision variables are only directly coupled to few other agent's variables through the objective function or the constraints. Nevertheless, existing algorithms for distributed optimization generally do not exploit the locality structure of the problem, requiring all agents to compute or exchange the full set of decision variables. In this paper, we develop a rigorous notion of ``locality" that quantifies the degree to which agents can compute their portion of the global solution based solely on information in their local neighborhood. This notion provides a theoretical basis for a rather simple algorithm in which agents individually solve a truncated sub-problem of the global problem, where the size of the sub-problem used depends on the locality of the problem, and the desired accuracy. Numerical results show that the proposed theoretical bounds are remarkably tight for well-conditioned problems.

% the structural properties of a linearly-constrained convex optimization problem (in particular, the sparsity structure of the constraint matrix and the objective function) to the amount of information that agents should exchange to compute an arbitrarily high-quality approximation to the problem from a cold-start. We leverage the notion of locality to develop a \emph{locality-aware} distributed optimization algorithm, and we show that, for problems where individual agents only require to know a small portion of the optimal solution, the algorithm requires very limited inter-agent communication. Numerical results show that the convergence rate of our algorithm is directly explained by the locality parameter proposed, and that the proposed theoretical bounds are remarkably tight for well-conditioned problems.

\end{abstract}
\section{Introduction}
Many problems in multi-agent control are naturally posed as large-scale optimization problems, where knowledge of the problem cost function and constraints is distributed among agents, and the collective actions of the network are summarized by a global vector decision variable. Concerns about communication overhead, privacy, and robustness in such settings have motivated the need for distributed solution algorithms that avoid explicitly gathering all of the problem data in one location. This is strikingly similar to a prominent setting in the literature on distributed optimization where knowledge of the objective function is distributed, i.e., can be expressed as the sum of privately known functions, and agents must reach a \emph{consensus} on the optimal decision variable despite limited inter-agent communication. We refer the reader to \cite{2017arXiv170908765N} for a recent survey on distributed optimization.

For many practical settings, seeking consensus as the end goal accurately represents the objective; for instance, in rendezvous and flocking problems, all the agents' actions depend on a global decision variable (meeting time and location for the former, and speed and heading for the latter). However, when the global decision variable represents a concatenation of individual actions, the network can still act optimally without ever coming to a consensus. Consider, for example, a task allocation problem where each agent only needs to know what tasks are assigned to itself, and is not concerned with other agents' assignments.

Many existing distributed optimization algorithms leverage consensus as a core building block and, broadly speaking, can be abstracted as the interleaving of descent steps, to drive the solution to the optimum, and averaging of information from neighbors, to enforce consistency. The main features differentiating these algorithms from each other are the centralized algorithm from which they are derived, and details regarding the communication structure such as synchronous or asynchronous, and directed or undirected communication links, with the broad overarching categories being consensus-based (sub)gradient (\cite{5404774}, \cite{6705625}), (sub)gradient push (\cite{1896df2670fa448e8f6ee55888ee65e7}, \cite{6120272}), dual-averaging (\cite{6426375}, \cite{Duchi.ea.TAC12}), and distributed second-order schemes (\cite{7134753}, \cite{7590162}). 

% \rbline{ }{actually, this entire following paragraph should better aligned with the statement of contribution. transition from previous paragraph by discussing literature on faster mixing via designing the laplacian. Q: since these existing algorithms leverage mixing over the whole network, what happens if we don't allow that. How well can we do with information only in a local neighborhood of ourselves.}

Historically, the mixing time of the communication graph has been seen as a fundamental limit on the convergence of distributed optimization algorithms \cite{Duchi.ea.TAC12}. Accordingly, a large body of the literature focuses on designing gossip matrices whose spectral properties allow for faster mixing of information \cite{Seidman.ea.CDC18}, \cite{Boyd.ea.06Gossip}. This perspective implicitly makes the assumption that convergence cannot be achieved until problem information has been disseminated and subsequently incorporated into the estimates of all of the agents. Our objective in this paper is to identify problems where this global mixing is an unnecessary overhead, by quantifying how well agents can compute their portion of the global solution based solely on information in their local neighborhood.

% \frmargin{Our main objective of this paper is to show that under reasonable assumptions about problem instances for multi-agent systems, the large communication overhead incurred by such algorithms is unnecessary.}{VERY strong - verify that it is supported by the text} Specifically, we will take advantage of sparsity structure in the constraints and objective to develop a notion of ``locality", namely the property that solution components can be computed with high accuracy without full knowledge of the problem. Under such assumptions, we will illustrate that an algorithm that restricts information exchange to ``where it matters most" is significantly more efficient than those which rely on eventually disseminating information through the entire network.

Our approach builds on the work of Rebeschini and Tatikonda~\cite{8370673}, who introduced a notion of ``correlation" among variables  %\frmargin{in purely-deterministic settings}{Clarify what purely deterministic means here. Do we include stochasticity? Or do we say this because ``correlation'' is typically used in stochastic settings? The sentence is ambiguous.} \rbmargin{as}{The latter. will figure out how to reword this so it doesn't seem like it's in contrast to wht we did} 
in network optimization problems. The authors in ~\cite{8370673} characterize the ``locality'' of network-flow problems, and show that the notion of locality can be applied to develop computationally-efficient algorithms for ``warm-start'' optimization, i.e., re-optimizing a problem when the problem is perturbed. Moallemi and Van Roy~\cite{5437447} have also explored similar notions of correlation, but solely as a tool to prove convergence of min-sum message passing algorithm for unconstrained convex optimization. To the best of our knowledge, \cite{8370673} is the only prior work to advocate for a general theory of locality in the context of multi-agent systems.

Our approach in this paper also draws influence from the field of local computation, a sub-field of theoretical computer science. Motivated by the common threads in problems such as locally decodable codes, reconstruction models, and decompression algorithms, Rubinfeld et al. \cite{DBLP:conf/innovations/RubinfeldTVX11} proposed a unifying framework of Local Computation Algorithms (LCAs). LCAs formalize the intuition that, in problems with large inputs and outputs, if only a small subset of the output is needed, it is inefficient to compute the entire output and simply read off the component required. Instead, both computation and access to the input should be kept to a minimum such that the required output is obtained and can be guaranteed to be consistent with subsequent queries. Kuhn et al. \cite{Kuhn:2016:LCL:2906142.2742012} adapted the definition of local computation for graph problems where nodes must make decisions based on information limited to a $k$-hop neighborhood around themselves. Under this model, they study the ``locality" of several prototypical combinatorial optimization problems and their linear programming relaxations, such as minimum vertex cover, and maximum independent set, by characterizing the bounds on estimation error as the $k$-hop neighborhood grows. 

% Orthogonal, yet complementary to the field of distributed optimization, local computation has recently emerged as a field in theoretical computer science.  However, due to the context in which LCAs were originally proposed, they do not consider the notion that inputs and outputs may have inherent \textit{geographic} locations.

\paragraph*{Statement of Contribution}
We develop a theoretical basis for the local-computation paradigm applied to convex optimization problems in multi-agent systems. Specifically, given the objective of computing $x^*_i$, a single component of the optimal decision variable, we characterize the error incurred by truncating the optimization problem to a neighborhood ``around" $x_i$. We show that for \emph{all} linearly-constrained strongly-convex optimization problems, this error decays exponentially with the size of the neighborhood at a rate dependent on the conditioning of the problem. This rate, which we \frmargin{coin as}{``name'' perhaps? Not a native english speaker, but this sounds weird. Feel free to disregard.} the ``locality" of a problem,  naturally characterizes the trade-off between the amount of local knowledge available to an agents, and the quality of its approximation. The condition number of a problem, colloquially referred to as a metric of how ``well-behaved" a problem is (where lower condition numbers are preferable), unsurprisingly, is positively correlated with the locality of a problem (where a low locality parameter indicates a rapid decay in influence of problem data). Our findings give a theoretical basis for a rather simple algorithm, in which agents simply solve truncated sub-problems of the global problem. Our numerical results, obtained by using this algorithm, show that the tightness of the theoretical bounds also depend on the condition number of the problem, with the bounds being near-optimal for well-conditioned problems.
% show that our bounds are remarkably tight for well-conditioned problems. 

A preliminary version of this work was accepted at the 2020 European Control Conference \cite{BrownRossiea20}. This paper extends prior results by providing tighter bounds on the locality of problems, in addition to showing that the property of locality is common to \emph{all} linearly-constrained strongly-convex optimization problems.

% \frmargin{A previous version of this paper was presented at the 2020 European Control Conference \cite{BrownRossiea20}. In this revised and extended version, we TODO.}{TODO I haven't read the full paper yet, but we should acknowledge similarities to previous work\rbmargin{.}{The technical overlap between this paper and the original ECC is the idea relating local sub-problems to the global problem, and sparsity of matrix powers part. I think if we tried to list the changes, it would be very confusing}}\ksmargin{-}{This description doesn't have to be exhaustive or super precise. For now, you can say that the current work provides a tighter / more elegant etc anaylsis, extended experimental work, and full proofs that were omitted in the conference version. }
% \ksmargin{-}{Building on top of Federico's remark, we should explicitly state how our work differs from previous work.}

\paragraph*{Organization}
In Section \ref{sec:notation}, we introduce notation, terminology, and technical assumptions about the problem. In Section \ref{sec:main_results}, we provide the problem statement, which establishes the fundamental question of locality, and summarize the main result, which establishes the ubiquity of locality in linearly-constrained convex optimization problems and provides a problem-specific bound on the rate of locality. We also summarize the key intermediary results, and discuss the algorithmic implications of locality in terms of the communication and message complexity it implies. Rigorous proofs of the main results are reported in Section \ref{sec:proofs}. In Section \ref{sec:experiments} we provide numerical experiments that highlight both scenarios where our theoretical bounds are tight, and those where our bounds are more conservative. Finally, we conclude and highlight future directions in Section \ref{sec:conclusion}.

\section{Notation and Assumptions}\label{sec:notation}
We use $[N]  := \{1,..., N\}$ to denote the $1 - N$ index set, and $e_i$ to denote the canonical $i$th basis vector i.e., the vector with $1$ in position $i$ and zero elsewhere, where the size of the vector will be clear from context. 
For a given matrix $A$, $A_{ij}$ denotes the element in the $i$th row and $j$th column of $A$. Similarly let $A_{i,*}$ and $A_{*,j}$ denote the  $i$th row and $j$th column of $A$ respectively. Let $A^T$ be the transpose, and $A^{-1}$ be the inverse of $A$. Given subsets $I \subseteq M, \, J \subseteq N$, let $A_{I,\,J} \in \mathbb{R}^{|I| \times |J|}$ denote the submatrix of $A$ corresponding to the rows and columns of $A$ indexed by $I$ and $J$, respectively. Similarly, let $A_{-I,\,-J}$ denote the submatrix of $A$  obtained by removing rows $I$ and columns $J$. We let $\sigma_\text{max}(A)$ and $\sigma_\text{min}(A)$ denote the maximum and minimum singular values of $A$ respectively, $\lambda_\text{max}(A)$ and $\lambda_\text{min}(A)$, the maximum and minimum eigenvalues, and $\kappa(A) = \frac{\vert\lambda_\text{max}(A)\vert}{\vert\lambda_\text{min}(A)\vert}$ the condition number. We define the difference between two sets, $S_1 \setdiff S_2 = \{s \in S_1 \mid s \not \in S_2\}$ to be the set of elements in $S_1$ but not $S_2$

Throughout this paper, we will consider linearly-constrained convex optimization problems of the form:
\begin{mini*}|l|
{x\in \mathbb{R}^N}{f(x) = \sum_i f_i(x_i)}
{}{}
\addConstraint{Ax = b.}\end{mini*}
We assume that $A \in \mathbb{R}^{M\times N}$ is full rank, and that each function $f_i: \mathbb{R}^N \rightarrow \mathbb{R}$ is $L$-smooth, $\mu$-strongly convex, and twice continuously differentiable. We let $V^{(p)} = [N]$  denote the set of primal variables, $V^{(d)}= [M]$ the set of dual variables, and  $S_j = \{i \in V^{(p)}| A_{ji} \neq 0\}$ the set of primal variables participating in the $j$th constraint. We also define the following set of constraints
\[C_{S \subseteq V^{(p)}} := \{i \in [M] \mid \text{ if } j \not \in S \text{ then } A_{ij} = 0 \},\]
% and \[\overline{C}_{S \subseteq V^{(p)}} := \{i \in [M] \mid A_{ij} \neq 0 \text{ for some } j \in S\}.\]
Intuitively, $C_S$ is the set of constraints that \emph{only} involve variables in $S$. Throughout this paper, we fix the objective function $f$ and the constraint matrix $A$, and write $x^*(b)$ as a function of the constraint vector, $b$.%, and $\overline{C}_S$ is the set of constraints that involve any of the variables in $S$. 

We define an undirected graph $G = (V,E)$ by its vertex set $V$ and edge set $E$, where elements $(i,j) \in E$ are unordered tuples with $i,j \in V$. We define the graph distance $d_G(i,j)$ to be the length of the shortest path between vertices $i$ and $j$ in graph $G$, and $\mathcal{N}^G_k(i) = \{j \in V \mid d_G(i, j) \leq k\}$ to be the $k$-hop neighborhood around vertex $i$ in graph $G$ for a given $k\in \mathbb{N}_{>0}$. We define the following undirected graphs representing coupling in the optimization problem:
\begin{itemize}
    \item  $G_\text{dec} = (V^{(p)}, \, E_\text{dec}(x))$, with $E_\text{dec} = \{(v^{(p)}_i,v^{(p)}_j)|A_{ki} \neq 0, \,  A_{kj} \neq 0\text{ for some } k \}$. Informally, $G_\text{dec}$ is the graph encoding the \emph{decision variables} that appear in the same constraint.
    \item $G_\text{con} = (V^{(d)}, \, E_\text{con})$, with $E_\text{con}= \{(i,j)| [AA^T]_{ij} \neq 0\}$. Informally, $G_\text{con}$ encodes connections between the \emph{constraints} through shared primal variables.
    \item  $G_\text{opt} = (V^{(p)} \cup V^{(d)}, \, E_\text{opt}(x))$, with $E_\text{opt} = \{(v^{(p)}_j,v^{(d)}_i)| A_{ij} \neq 0\}$. Informally, $G_\text{opt}$ is the graph encoding the overall dependence structure of the \emph{optimization problem}.
\end{itemize}

\section{Foundations of locality, and their algorithmic implications}\label{sec:main_results}
\subsection{Problem Statement}
\label{subsec:setup}
%\rbmargin{*}{	- I would like this section to read as: Problem - Does the most naive thing work? Main Result - Yes. Here is how well. Algorithmic Implication - Directly applying to main results helps with the problem of communication (which was how we motivated solving this problem)}
We consider a network of $N$  agents collectively solving the following linearly-constrained optimization problem
\begin{mini}|l|
{x\in \mathbb{R}^N}{f(x) = \sum_i f_i(x_i)}
{\label{global_opt}}{}
\addConstraint{Ax = b,}
\end{mini}
where knowledge of the constraints is distributed, and the decision variable represents a concatenation of the decisions of individual agents. Specifically, we assume that $f_j$ and $A_{*j}$ are initially known by agent $j$ only, and agent $j$ knows $b_i$ if $A_{ij} \neq 0$. As a motivating example, consider a scenario where a fleet of agents needs to collectively complete tasks at various locations, while minimizing the cost of completing such tasks. In this setting, the constraints ensure completion of the tasks, while the entries $A_{ij}$ of the constraint matrix may encode the portion of task $i$ that agent $j$ can complete, or efficiency when completing tasks, thus, constituting private knowledge. See Section~\ref{sec:experiments} for additional examples. 

As a departure from a large body of the existing literature on distributed optimization, we consider the problem to be solved when each agent $j$ knows $x^*_j$---that is, we do not require every agent to know the entire optimal decision variable. With some abuse of notation, we conflate each agent with its associated primal variable\footnote{While, in this paper, each agent is only associated with a scalar variable for illustrative purposes, one can readily extend the results in this paper to the setting where each agent is associated with a vector. Additionally, the case where multiple agents' actions depend on shared variables can be addressed by creating local copies of those variables and enforcing consistency between agents who share that variable through a coupling constraint.}.

Our objective in this paper is to characterize the accuracy with which an agent $i$ can compute its associated solution $x_i$ component given access to problem data held by agents within a $k$-hop neighborhood of itself in $G_\text{dec}$, for a given $k\in \mathbb{N}_{>0}$. On the communication graph given by  $G_\text{dec}$, obtaining this information requires $k$ communication rounds of accumulating and passing problem data between neighbors. Consequently, our results also characterize the trade-off between communication and approximation accuracy in this setting. This communication graph should not be seen \ksmargin{as prescriptive}{I don't understand this. can you elaborate?}, but rather one that facilitates ready analysis of locality in multi-agent systems.

\subsection{Foundations of Locality}
For each $x_i$, we consider sub-problems induced by restricting Problem \eqref{global_opt} to the variables within the $k$-hop neighborhood around $x_i$ and constraints only involving those variables (termed ``$k$-hop local sub-problems"). The main result of this paper states that the error in the $i$th component of the $k$-hop ``local solution" decays exponentially with the size of the neighborhood. A formal statement of this result is provided below.
\begin{theorem}[Locality]\label{thm:locality_sketch}
Let $x^{(k)}$ be the solution to the optimization problem induced by restricting Problem \ref{global_opt} to $k$-hop neighborhood around $x_i$,  $\mathcal{N}_k^{\text{(dec)}}(i)$, and the constraints only involving those variables. If \ksmargin{$\lambda = \sup_x \frac{\sqrt{\kappa(x)} - 1}{\sqrt{\kappa(x)} + 1}$} {you may want to state $\lambda$ in a preceding standalone definition, and call with a fancy name, e.g., locality rate.}, where $\kappa(x)$ denotes the \ksmargin{condition number}{unless this is standard, it may be worth providing a definition for this term, or for the very least reference a source where it's defined} of $A \nabla^2 f(x)^{-1} A^T$, then
\begin{equation}
\lvert x^{(k)}_i - x^*_i \rvert \leq C \lambda^k
\label{eq:bound_x}
\end{equation}
for $C = \left(1+ \sqrt{\frac{L}{\mu}}\right)\frac{2\sigma_{max}(A)}{\sigma^2_{min}(A)}\norm{ b - Ax^*_{UC}}_2$.
\end{theorem}
The rate $\lambda$ naturally characterizes the degree to which local information is sufficient to approximate individual components of the global optimum, thus justifying it as a metric of ``locality". The proof of Theorem \ref{thm:locality_sketch} relies on two intermediary results.

Our first intermediary result derives the relationship between solutions to the local sub-problems and the true solution to Problem \eqref{global_opt} (the ``global problem"). Specifically, we show that the solution to a local sub-problem is consistent with that of a perturbed version of the global problem (where the perturbation appears in the constraint vector, $b$).
\begin{theorem}[Relationship between local sub-problems and the global problem]\label{thm:loc_global_sketch}
Let $S\subseteq V^{(p)}$ be a subset of the primal variables. If $x^{(S)}$ is the solution to the problem obtained by restricting Problem \eqref{global_opt} to the variables in $S$ and constraints only involving those variables, i.e., 
\begin{argmini}|l|
{x^{(S)}\in \mathbb{R}^{|S|}}{\sum_{i \in S} f_i\left(x^{(    S)}_i\right),}
{}{x^{(S)} = }
\addConstraint{A_{C_s, S} x^{(S)} = b_{C_s},}
\end{argmini}
then there exists $\hat{b} \in \mathbb{R}^M$ such that $x^{(S)}= \left[x^*(\hat{b})\right]_S$.
\begin{proof}[Proof Sketch]
We proceed by showing that augmenting the local sub-problem with the remaining variables does not change the solution on the local-subproblem. By computing the values that the remaining constraints naturally take on (without being enforced), we can derive the global constraint vector $\hat{b}$ that induces the same value on $S$.
\end{proof}
\end{theorem}

The importance of Theorem \ref{thm:loc_global_sketch} lies in the fact that  we can interpret solving local sub-problems as solving perturbed versions of the global problem. This interpretation allows us to leverage theory on the sensitivity of optimal points of Problem \eqref{global_opt} to characterize the error incurred by only using a subset of the original problem data.

Our second intermediary result characterizes the component-wise magnitudes of this correction factor. Specifically, we show that when the constraint vector of Problem \ref{global_opt} is perturbed, the impact of the perturbation decays exponentially with distance to the perturbation.
\begin{theorem}[Decay in sensitivity of optimal points]\label{thm:decay_sketch}
Let $\lambda$ be defined as in Theorem \ref{thm:locality_sketch}. Then for any perturbation in the constraint vector, $\Delta \in \mathbb{R}^M$, subset of the primal variables, $S \subseteq V^{(p)}$, and $C = \frac{2\norm{\Delta}_2}{\sigma_\text{min}(A)}$,
\[\left\|\left[x^*(b + \Delta) - x^*(b)\right]_S\right\|_2 \leq C \lambda^{d(S, \text{supp}(\Delta))}.\]
\begin{proof}[Proof Sketch]
The proof leverages the Conjugate Residuals algorithm (designed to solve linear systems) to generate a sequence of sparse approximations to the correction factor that converges exponentially to the true correction factor. The sparsity patterns of the approximate correction factors reflect the underlying graph structure of the optimization problem, and can be determined \emph{a priori}. The convergence guarantees of conjugate residuals along with a characterization of the sparsity pattern of its iterates allow us to derive a component-wise bound on the correction factor. 
\end{proof}
\end{theorem}
Intuitively, this theorem states that a perturbation in the constraints affects the decision variables ``closest" to the constraint the most, i.e., those that are actually involved in the constraint, while the effect of the perturbation decays with the degrees of separation between a decision variable and the constraint. The construction of the $k$-hop local sub-problems takes advantage of this theorem by forcing the ``perturbation" to be at a distance of at least $k$ from component $x_i$. Theorem \ref{thm:locality_sketch} is derived from the intermediary results by bounding the perturbations induced by cutting constraints.
\subsection{Algorithmic Implications}
% \section{Exploiting Locality for Distributed Optimization}
% \label{subsec:algorithm}
%\rbline{ }{This section no longer needs to introduce the idea of cutting constraints/generating sub problems. Those notion are already introduced in the previous section, and are inherent to the definition of locality. Instead, this section should make clear the connection between communication volume and size of subproblem. Maybe present in the context where $G_{dec}$ is the communication graph but show that it is up to a constant factor different when on a different comms graph. Also should acknowledge the weakness of message size}
\frmargin{The characterization}{Is this section ``algorithmic implications'' or ``a locality-aware optimization algorithm''? Also, mention algorithm in statement of contribution.} of locality naturally suggests a means of reducing the communication necessary for distributed optimization. In a radical departure from much of the existing work on distributed optimization, which rely on propagating information throughout the network, we suggest \emph{localizing} information flow. Our results show that the importance of problem data to individual solution components decays with distance to the data. Consequently, if a problem exhibits sufficient locality, by restricting information flow to where it matters most, we can avoid the high communication overhead of flooding methods with little impact on solution quality. 

% In addition to the assumptions of Section \ref{subsec:setup}, we will now assume there is a bidirectional communication link between agents $i$ and $j$ if $A_{ki} \neq 0$ and $A_{kj} \neq 0$ for some $k$, i.e., if agents $i$ and $j$ participate in the same constraints. Finally, we assume that each local function can be fully represented by $B$ bits. 

The objective is for each agent to compute its own component of the solution vector, i.e., for agent $i$ to compute $x^*_i$. 
We denote by $\hat{x}_i$ agent $i$'s estimate of $x^*_i$ and we let $\hat{x} = (\hat{x}_1, \ldots,  \hat{x}_N)$ be the aggregation of privately known solution components. Because we allow the approximation to violate constraints, the typical metric of sub-optimality in the objective function is \ksmargin{uninformative}{can you clarify?}---the approximation generated is guaranteed to have an  objective value no larger than the true optimum. Consequently, we will measure the accuracy of our solution by $\norm{\hat{x} - x_*}_\infty$---this bound readily translates into bounds on both the objective value and constraint violation if such metrics are preferred. We note that by strong convexity of the objective, the optimal solution is guaranteed to be unique. This ensures that our notion of an approximate solution is well-defined, and rules out the case of ``jumps" to other optimal solutions.

\frmargin{The locality-aware distributed optimization algorithm}{This is the first time we talk about "the" locality-aware distributed optimization algorithm, save for the abstract. We should restructure this section a bit.} is conceptually simple. Leveraging locality, we conclude that each agent can compute its component of the solution by solving a local sub-problem of the global problem, where the size of the local sub-problem depends on the accuracy desired and the locality parameter of the global problem. Agents aggregate local problem data through a recursive flooding scheme, which is truncated after a predetermined number of communication rounds. Then, each agent solves its own local problem without further communication with the network.
Specifically, agent $i$ starts with its local objective function, $f_i$, its associated column of the constraint matrix $A_{*i}$, and components of the constraint vector $b_{C_i}$. In the initialization phase, agent $i$ sends $A_{C_ii}$ to each of its neighbors. After the initialization phase, agent $i$ has full knowledge of $A_{C_i *}$, i.e., the constraints that it participates in. Then, in the first iteration, agent $i$ sends a representation of $A_{C_i*}$, $b_{C_i}$ and $f_i$ to each of its neighbors. In subsequent iterations, each agent sends a representation of all of the information it has previously received to each of its neighbors. After the $k$'th iteration, for $k\in [K]$, agent $i$ has a representation of $f_j$, $b_{C_j}$ and $A_{C_j*}$ for all $j \in \mathcal{N}(i, k)$, where $\mathcal{N}(i, k)$ denotes the $k$-hop neighbors of agent $i$. After the $K$ communication rounds, agent $i$ generates its local sub-problem by ignoring any constraints involving variable outside of its $K$-hop neighborhood, $\mathcal{N}(i, K)$.
The algorithm for agent $i$ is summarized in Algorithm \ref{alg:locality}. 

\begin{algorithm}[!ht]
\SetKwInOut{Input}{input}
\Input{$f_i, \, A_{* i}, \, b_{C_i}, \, K$}
Initialization: Send $A_{C_i i}$ to all $j \in \mathcal{N}(i, 1)$\;
\For{$k = 1, \ldots, K$}{
Send $\{f_l, \, A_{C_l, *}, \, b_{C_l}\}_{l \in \mathcal{N}(i, k - 1)}$ to all $j \in \mathcal{N}(i, 1)$\;
}
Solve 
\nonl \vspace{-\baselineskip} \begin{argmini}|s|
{x\in \mathbb{R}^{|\mathcal{N}(i, K)|}}{\sum_{j \in \mathcal{N}(i, K)} f_j(x_j)}
{}{x^{(\mathcal{N}(i, K))} = }
\addConstraint{A_{C_{\mathcal{N}(i, K)}}x = b_{C_{\mathcal{N}(i, K)}}}
\end{argmini}
% where $C := \{ c \in C_{\mathcal{N}(i, K)} \mid A_{cj} \neq 0 \text{ and} j\not\in \mathcal{N}(i, K)\}$\;
$\hat{x_i} = x_i^{(\mathcal{N}(i, K))}$
\caption{Locality-Aware Distributed Optimization}
\label{alg:locality}
\end{algorithm}
\subsection{Discussion}\label{subsec:Discussion}
It follows directly from the locality analysis in Theorem~\ref{thm:locality_sketch} that if an accuracy of $\norm{ \hat{x} - x_*}_\infty \leq \epsilon$ is desired, \[K \geq \frac{1}{1-\lambda}\log\left(\frac{C}{\epsilon}\right)\] communication rounds are sufficient. This bound not only determines how to select the number of communication rounds (passed in as a hyperparameter to the algorithm), but should be seen as guidance in determining whether the locality-aware algorithm is suitable for a particular setting. If $K$ is greater than the radius of the network, at least one node has accumulated the entirety of the problem data, and if $K$ is greater than the diameter of the network, the locality-aware algorithm amounts to accumulating and solving the entirety of the problem data at every node---in such settings, the locality-aware algorithm may not be suitable. Generally, the locality-aware algorithm offers an advantage in scenarios where the locality parameter, $\lambda$, is sufficiently small, and the network diameter is much larger than $K$.
%\ksmargin{-}{end this with a more positive tone so that the reader won't get the impression that this approach is useless. }

In contrast to algorithms where estimates of the primal or dual solutions are passed between agents, the message complexity of the proposed algorithm is not constant across iterations---the size of the messages grows at each iteration with the number of agents in each expanding neighborhood. Explicitly, if each local function can be fully represented by $B$ bits, a message representing $\{f_i, \, A_{C_i,*} , \, b_{C_i}\}$ requires on the order of $\mathcal{O}(B + 4 \max_i{|S_i|}\times\max_j{|C_j|})$ bits, where $\max_i{|S_i|}$ is the maximum number of agents participating in a constraint, and $\max_j{|C_j|}$ is the maximum number of constraints any agent participates in. Because $|\mathcal{N}(i, k - 1)| \leq \left(\max_i{|S_i|}\times\max_j{|C_j|}\right)^{k-1}$, the maximum message size during the $k$th communication round is on the order of $\mathcal{O}\left(\left(\max_i{|S_i|}\times\max_j{|C_j|}\right)^{k}\right)$ bits. %Consequently, the size of the message sent at iteration $k$ by $i$ to each of its neighbors is given by $|\mathcal{N}(i, k - 1)|(B + 4 \max_i{|S_i|}\max_j{|C_j|})$. By definition, $|\mathcal{N}(i, 1)| \leq \max_i{|S_i|}\times\max_j{|C_j|}$, so it holds true that $|\mathcal{N}(i, k - 1)| \leq \left(\max_i{|S_i|}\times\max_j{|C_j|}\right)^{k-1}$. This shows that the maximum message size during the $k$th communication round is on the order of \[\left(\max_i{|S_i|}\times\max_j{|C_j|}\right)^{k}\] bits. 

Notably, both the number of communication rounds and the message complexity of the locality-aware algorithm do not directly depend on the number of nodes in the network. In contrast, distributed optimization algorithms that rely on passing estimates of the decision variable requires each node to send messages of size $\mathcal{O}(N)$ at every iteration. Moreover, the number of iterations to convergence of such methods tend to scale with the number of nodes in the network (depending on network topology) \cite{2017arXiv170908765N}.\todo{address partial consensus} While the message complexity of the locality-aware algorithm grows rapidly between iterations, when $A$ is sparse, $|S_i| \ll N$ and $|C_i| \ll M$. This analysis suggests that the locality-aware algorithm offers a significant advantage in settings where $|S_i|$ and $|C_i|$ remain bounded as $N$ and $M$ are increased, i.e., those where a bounded number of agents participate in constraints, and agents participate in a bounded number of constraints regardless of the size of the network. 

%\frmargin{}{There is no direct dependency, true. But, for the same problem, does $\lambda$ change with the problem size? If not, mention. If yes, maybe point to numerical experiments.}. 
A shortcoming of Algorithm \ref{alg:locality} is that problem data is explicitly shared between agents. At present, its application is limited to settings where preserving the privacy of individual objective functions and constraint sets is not a concern. However, the scalability of the locality-sensitive algorithm in such settings motivates extending these ideas to design algorithms that exploit locality without explicitly sharing problem data, and we highlight as a promising future direction. 

\section{Proofs of Main Results}\label{sec:proofs}
% In this section, we propose a rigorous metric of the ``locality" of an optimization problem---namely, a measure of the extent to which individual components of the decision variable can be solved for with reasonable accuracy using only a localized subset of the original problem data. Our approach to this problem is twofold.
In this section, we prove the main results summarized in Section \ref{sec:main_results}. First, in Section \ref{subsec:local_global}, we derive the relationship between the true solution to Problem \eqref{global_opt} (termed the ``global problem") and the solution to the problem obtained  by restricting Problem \eqref{global_opt} to a subset of the variables and the constraints only involving those variables (termed the ``local sub-problem"). Explicitly, we show that the solution of the local sub-problem is consistent with the solution of a perturbed version of the global problem. This then allows us to leverage the sensitivity expression in \cite{8370673} to derive an expression for the difference between the solution to the local sub-problem and the solution to the global problem (henceforth denoted as the ``correction factor").% \frmargin{determine the correction factor that will drive the solution of the local subproblem to that of the global problem}{YOu never define ``correction factor''. I suggest "}. 

Second, in Section \ref{subsec:sensitivity}, we show that the correction factor derived in Section \ref{subsec:local_global} yields a numerical structure that reflects the underlying structure of the constraints. Specifically, we show that, while the correction factor will typically be dense, it is possible to find a sparse approximation to the correction factor, where the sparsity pattern of the approximation is a function of the sparsity of the constraints, the desired accuracy, and the conditioning of the global problem. We leverage the guarantees of the Conjugate Residual algorithm to derive, \emph{a priori}, both the sparsity pattern and a bound on the accuracy of the approximation. This approach will allow us to identify which elements of a local solution will be unaffected if a sparse approximation of the correction factor is used.
%\frmargin{(or, equivalently, the amount of communication between agents)}{Robin, is this correct?}
In Section \ref{subsec:summary}, we use the results of the previous sub-sections to characterize the relationship between the quantity of problem data used, and the error in individual components. This will naturally give rise to the metric of locality $\lambda$, which we formally present at the end of the section.
\subsection{Relating local sub-problems to the global problem}\label{subsec:local_global}
In this section, we consider sub-problems generated by restricting Problem \eqref{global_opt} to a subset of the primal variables and the constraints only involving those variables. In particular, if $S\subseteq V^{(p)}$ is a subset of the primal variables, we define \[C_{S \subseteq V^{(p)}} := \{i \in [M] \mid \text{ if } j \not \in S \text{ then } A_{ij} = 0 \}.\] These are the constraints of the global problem that \emph{only} involve the components of $S$. We define the following problem to be the \emph{local sub-problem induced by $S$}:
\begin{argmini}|l|
{x^{(S)}\in \mathbb{R}^{|S|}}{\sum_{i \in S} f_i(x^{(S)}_i)}
{\label{eq:loc_opt}}{x^{(S)} = }
\addConstraint{A_{C_s, S} x^{(S)} = b_{C_s}.}
\end{argmini}
Our objective in this section is to relate the value of $x^{(S)}$ to $\left[x^*(b)\right]_S$, the components $S$ of the global optimum. Ultimately, this will allow us to analyze the error incurred by only using a subset of the original problem data. 

We first show that augmenting the local sub-problem with the remaining variables does not change the solution to the local sub-problem. By computing the values that the cut constraints naturally take on (without being enforced), we can derive the global constraint vector $\hat{b}$ that induces the same value on $S$, i.e., $x^{(S)} = \left[x^*(\hat{b})\right]_S$. This equivalence allows us to exploit the sensitivity of optimal points of Problem \eqref{global_opt} to perturbations in the constraint vector, $b$, to derive the correction factor that drives the solution of the local sub-problem to that of the global problem. This interpretation is key for making the connection between the ``warm-start" scenario presented in \cite{8370673} (where the algorithm needs to compute $x^*(b)$ given the solution to $x^*(b + p)$)  to the ``cold-start" scenario considered in this paper (where the algorithm must compute $x^*(b)$ without prior knowledge of other optimal solutions). The allows us to develop a more general theory of locality that fully captures the importance of problem data to individual solution components, as opposed to a theory that only captures response to perturbations. 
%\rbline{ }{This whole following section is stupid. In the proof of the main theorem, we just augment the local function with the remaining unconstrained variables. There's no point in introducing the idea of cutting constraints...Only the implicit constraints part is important. On the bright side, I can cut a lot of length here}

In the following lemma, we show that if the local-sub-problems are augmented with the remaining variables, the solution on the $k$-hop neighborhood does not change. 
\begin{lemma}\label{lem:loc_ext}
Let $x^{(S)}$ be the solution to the local sub-problem induced by $S$, and 
\begin{argmini}|l|
{x\in \mathbb{R}^{N}}{\sum_{i = 1}^N f_i(x_i)}
{\label{eq:loc_est_opt}}{\hat{x}^{(S)} = }
\addConstraint{A_{C_s, S} x = b_{C_s}.}
\end{argmini}
is the solution to the problem including the entire objective function, but only the constraints of the local sub-problem, then $x^{(S)} = \left[\hat{x}^{(S)}\right]_S$.
\begin{proof}This lemma follows from observing that the variables in $V^{(p)} \setdiff S$ are entirely unconstrained, and can be optimized independently from those in $S$.
\end{proof}
\end{lemma}

By computing the values that the constraints in $V^{(d)} \setdiff C_s$ take on without being enforced, we can derive a constraint vector $\hat{b}$ that induces the same optimal solution as the partitioned problem (termed the ``implicit constraints"). The following Lemma formalizes this.
\begin{lemma}[Implicit Constraints]\label{lem:implicit_constraints}
Let $\hat{x}^{(S)}$ be defined as in Lemma \ref{lem:loc_ext}, and $\hat{b} = A \hat{x}^{(S)}$. Then, 
\begin{argmini}|l|
{x\in \mathbb{R}^N}{f(x)}
{\label{eq:implicit}}{\hat{x}^{(S)} = }
\addConstraint{Ax = \hat{b}}.
\end{argmini}
\begin{proof}
Assume by contradiction that there exists an optimal solution $\tilde{x}^*\neq \hat{x}^{(S)}$ to Problem \eqref{eq:implicit} with optimal value ${f(\tilde{x}^*)<f(\hat{x}^{(S)})}$.
Note that on $V^{(d)} \setminus C$, the implicit constraints are equal to the true constraints. Precisely,  $b_{C_s} = \left[\hat{b}\right]_{C_s}$. 

The constraints in Problem \eqref{eq:loc_est_opt} are a subset of the constraints in Problem \eqref{eq:implicit}. Therefore, the feasible set of Problem \eqref{eq:implicit} is contained in the feasible set of Problem \eqref{eq:loc_est_opt}. Explicitly,
\begin{align*}\{x \mid Ax = \hat{b}\}  &=  \{x \mid A_{-C,*}x = \hat{b}_{-C},\, A_{C,*}x = \hat{b}_C\}\\  &\subseteq \{x \mid A_{C,*}x = \hat{b}_{C}\}.\end{align*}
Therefore, if $\tilde{x}^*$ is the optimal solution to Problem \eqref{eq:implicit}, it is also a feasible solution for Problem \eqref{eq:loc_est_opt}. Since $f(\tilde{x}^*)<f(\hat{x}^{(S)})$, $\hat{x}^{(S)}$ is not optimal for Problem \eqref{eq:loc_est_opt}---a contradiction.
\end{proof}
\end{lemma}

% Lemma \ref{lem:partition} equates solving the local sub-problem induced by $S$ with solving the global problem where $b$ is replaced by $\hat{b}$. This equivalence allows us to use sensitivity of the 

Lemma \ref{lem:implicit_constraints} allows us interpret solving the local sub-problem as solving a perturbed version of the global problem where $b$ is replaced by $\hat{b}$. This interpretation allows us to leverage the theory developed by Rebeschini and Tatikonda \cite{8370673} on the sensitivity of optimal points of Problem \eqref{global_opt} to finite perturbations in the constraint vector, $b$, to relate the solution of the local sub-problem to that of the global problem. The main theorem of \cite{8370673} is reviewed below.
%\frmargin{derive the correction factor that drives the solution of the local sub-problem to that of the global problem}{"characterize the error between the solution of the local sub-problem and the solution to the global problem." I suspect you think of it as a "correction factor" because you integrate through a path. But either you define the correction factor as $x^*-\hat x^*$ (like you do later in the paper) or you define it as the thing you integrate. Either way, it should be defined clearly.}.

\begin{theorem}[Sensitivity of Optimal Points - Theorem~1 of \cite{8370673}]\label{RT_1}
Let $f: \mathbb{R}^N \rightarrow \mathbb{R}$ be strongly convex and twice continuously differentiable, and $A\in \mathbb{R}^{M\times N}$ have full row rank. For $b\in \text{Im}(A)$, let $\Sigma(x^*(b)) := \nabla^2 f(x^*(b))^{-1}$ . Then $x^*(b)$ is continuously differentiable at all $b \in \mathbb{R}^m$, and 
% $A\Sigma(x^*(b))A^T$ is invertible, and \frmargin{ }{This seems grammatically disconnected from the rest of the statement. What does this mean?}, and 
\begin{equation}
\frac{\mathrm{d}\,x^*(b)}{\mathrm{d}\,b} = D(b) =  \Sigma(x^*(b))A^T(A\Sigma(x^*(b))A^T)^{-1}. \label{sensitivity}
\end{equation}
\end{theorem}
The above theorem relates the gradient of the optimal solution, $x^*(b)$, to the constraint matrix and the objective function. Critically, Equation \eqref{sensitivity} holds globally, allowing us to apply the Fundamental Theorem of Calculus \frmargin{to determine the correction factor necessary to correct}{here you define correction factor = $x^* - \hat{x}^*$. Intuitively, ``error'' is a better term for this. See comments above.} for finite perturbations in the constraint vector.  Precisely, if we let  $\Delta = b - \hat{b}$, the correction factor can be expressed as
\begin{equation*}
x^*(\hat{b} + \Delta) - x^*(\hat{b}) =  \left(\int_0^1 \Sigma(x_\theta)A^T(A\Sigma(x_\theta)A^T)^{-1} d\theta\right)\Delta,
\end{equation*}
%is based on Hadamard's global inverse function theorem (rather than the implicit function theorem, which holds only locally). The significance is that the derivative of the optimal point is continuous \emph{everywhere} along the subspace $\text{Im}(A) = \mathbb{R}^M$. This allows 
% \begin{align*}
%      x^*(\hat{b} + \Delta) - x^*(\hat{b}) &= \left(\int_0^1 \frac{dx^*(\hat{b} + \theta\Delta)}{d\theta} d\theta\right)\Delta\\
%     &= \left(\int_0^1 \Sigma(x_\theta)A^T(A\Sigma(x_\theta)A^T)^{-1} d\theta\right)\Delta,
% \end{align*}
where $x_\theta := x^*(\hat{b} + \theta\Delta)$.
Consequently, we have established that the error incurred by only using a subset of the original problem data is precisely this correction factor. %\frmargin{An equivalent, and more suggestive, interpretation is  }{I think I am not following your intuition properly. To me, the two interpretations are identical, and both come down to "this is the difference between what I get and what I would like to get, i.e. the error". Am I missing some nuance?}
\subsection{Component-wise Sensitivity}\label{subsec:sensitivity}
% Consider a scenario where the solution to $x^*(\hat{b})$ is known, and we perturb the constraint vector by $\Delta$. Then, the solution to the perturbed problem can be computed as \[x^*(\hat{b} + \Delta) = x^*(\hat{b}) + \left(\int_0^1 \Sigma(x_\theta)A^T(A\Sigma(x_\theta)A^T)^{-1} d\theta\right)\Delta,\] the sum of the unperturbed solution and a correction factor.  
In the previous section, we gave a closed-form expression for the error incurred by not using the entire problem data. In this section, we show how the underlying structure of the optimization problem is reflected in the numerical structure of the correction factor. In particular, we leverage the Conjugate Residuals algorithm \cite{Saad03} (designed to solve linear systems) to generate a sequence of sparse approximations to the correction factor that converge exponentially to the true correction factor while maintaining sparsity patterns that reflect the underlying graph structure of the optimization problem. We establish that a perturbation in the constraints affects the decision variables ``closest" to the constraint the most, i.e., those that are actually involved in the constraint, while the effect of the perturbation decays with the degrees of separation between a decision variable and the constraint. Moreover, we derive an \emph{a priori} bound of the rate of decay.

In the remainder of this section, we will analyze the instantaneous sensitivity of the optimal point 
\begin{equation*}
\frac{\mathrm{d}\,x^*(b)}{\mathrm{d}\,b}\Delta = D(b)\Delta = \Sigma(x^*(b))A^T(A\Sigma(x^*(b))A^T)^{-1}\Delta.
\end{equation*}
In Section \ref{subsec:summary}, when we formally define our metric of locality, the results developed in this section will naturally extend to finite perturbations in the constraint vector. For ease of notation, we let $\Sigma = \Sigma(x^*(b))$. 

The instantaneous sensitivity expression will allow us to reason about the structural coupling between components of Problem \eqref{global_opt}, however, the term $(A \Sigma A^T)^{-1}$ will require careful treatment. Specifically, the inverse of sparse matrices is not guaranteed to be sparse, and in fact, is typically dense. While the structure of $A \Sigma A^T$ is obfuscated when we take the inverse, it is not lost. The insight that allows us to recover the original structure of the problem in the sensitivity expression is that the Conjugate Residuals algorithm can be leveraged to generate structure-preserving sparse approximations to $\delta := (A \Sigma A^T)^{-1} \Delta$. We now provide a high-level overview of the algorithm and relevant guarantees \cite[6.8]{Saad03}\footnote{We adapt the results from \cite{Saad03} slightly because $A\Sigma A^T$ is known to be normal.}.

\paragraph{Conjugate Residuals} For ease of notation, let $M = A \Sigma A^T$. Conjugate residuals (CR) is an iterative Krylov method for generating solutions to linear systems, $\delta = M \Delta$, where $M$ is a symmetric positive definite matrix. Specifically, the algorithm recursively generates a sequence of iterates \[\delta^{(k)} \in \mathcal{K}(M, \Delta, k) := \text{span}\{\Delta, \, M\Delta, \, M^2 \Delta, \ldots, M^{k - 1} \Delta \}\]
where each $\delta^{(k)}$ minimizes the norm of the residuals, $\norm{r_k} := \norm{\Delta - M\delta^{(k)}}_2$, in the $k$th Krylov subspace. The guarantees of the algorithm that we will leverage are as follows,
\begin{enumerate}
    \item  Sparsity: \\ $\delta^{(k)} \in \mathcal{K}(M, \Delta, k) := \text{span}\{\Delta, \, M\Delta, \, M^2 \Delta, \ldots, M^{k - 1} \Delta \}$.
    % \item \begin{align*}
    %     \norm{\delta^{(k)} - \delta}_M &\leq 2\left(\dfrac{\sqrt{\kappa} - 1}{\sqrt{\kappa} + 1}\right)^k \norm{\delta^{(0)} - \delta}_M\\
    %     &= 2\left(\dfrac{\sqrt{\kappa} - 1}{\sqrt{\kappa} + 1}\right)^k \norm{\delta}_M\\
    %     \norm{\delta^{(k)} - \delta}_2 &\leq 2\sqrt{\kappa} \left(\dfrac{\sqrt{\kappa} - 1}{\sqrt{\kappa} + 1}\right)^k \norm{\delta}_2
    % \end{align*}  (Convergence Rate)
    \item Convergence Rate: \begin{align*}
        \norm{r_k}_2 &\leq 2\left(\frac{\sqrt{\kappa} - 1}{\sqrt{\kappa} + 1}\right)^k \norm{r_0}_2 = 2\left(\frac{\sqrt{\kappa} - 1}{\sqrt{\kappa} + 1}\right)^k \norm{\Delta}_2.
    \end{align*} 
    %\item Finite time solution: $\norm{r_k}_2 = 0$ for some $k \leq N$
\end{enumerate}
The first guarantee will allow us to derive the support of each $\delta^{(k)}$, which reflects the underlying structure of the global problem. The second guarantee will allow us to prove the rate with which the effect of a perturbation decays with each degree of separation.

% We now define several graphs that will facilitate reasoning about the sparsity pattern of each of the approximations. This in turn will allow us to deduce the numerical structure of the sensitivity expression. We define the following undirected graphs:
% \begin{itemize}
%     \item $G_\text{con} = (V^{(d)}, \, E_\text{con})$, with $E_\text{con}= \{(i,j)| [AA^T]_{ij} \neq 0\}$. Informally, $G_\text{con}$ encodes connections between dual variables through shared primal variables.
%     \item $G_\text{opt} = (V^{(p)} \cup V^{(d)}, \, E_\text{opt}(x))$, with $E_\text{opt} = \{(v^{(p)}_j,v^{(d)}_i)| A_{ij} \neq 0\}$. Informally, $G_\text{opt}$ is the bipartite graph showing the primal variables involved in each constraint.
% \end{itemize}
% Using these graphs, the following theorem and corollary will allow us to derive the sparsity pattern of each approximation $\delta^{(k)}$. This, in turn, can be used to derive the sparsity pattern when they are pre-multiplied by $\Sigma A^T$.
\paragraph{Support of the estimates}
\begin{theorem}[Sparsity Structure of Matrix Powers]\label{thm:sens_supp}
For $k \in \mathbb{Z}_+$, neglecting numerical cancellation\footnote{When characterizing the sparsity pattern of a matrix, ``numerical cancellation" refers to entries that are zeroed out due to the exact values of entries in the matrix, and cannot be deduced to be zero from the combinatorial structure of the matrix alone.},  
\begin{equation*}
\text{supp}((A\Sigma A^T)^k) = \{(i,j)\mid d_{G_\text{con}}(v_i, v_j) \leq k\}.
\end{equation*}
% \begin{proof}
% \rbline{ }{TODO}
% \end{proof}
\end{theorem}
This theorem establishes that the sparsity pattern of a symmetric matrix to the $k$th power is determined by the $k$-hop neighbors in the graph representing the sparsity pattern of the original matrix. This allows us the characterize the sparsity pattern of each of the generating vectors of the $k$th Krylov subspace generated by $A\Sigma A^T$ and $\Delta$. 
%We define the set $\mathcal{N}^{(d)}_k (i):=\{v\in V^{(d)}| d_{G_\text{con}}(v^{(d)}_i, v) \leq k\}$ to be the set of vertices of distance at most $k$ from vertex~$i$ in $G_\text{con}$.
\begin{corollary}[Sparsity Structure of the Sensitivity Expression]\label{cor:sens_distr} For $k \in \mathbb{Z}_+$ and $i \in [M]$
\[ \text{supp}\left(\Sigma(x)A^T\delta^{(k)}e_i\right)\\ \subseteq \mathcal{N}_1^{G_\text{opt}}(\mathcal{N}^{G_\text{con}}_{k - 1}(i)).\]
\end{corollary}

Informally, $\mathcal{N}_1^{G_\text{opt}}(\mathcal{N}^{G_\text{con}}_{k - 1}(i))$ represents the components of $\Sigma A^T\delta^{(k)}e_i$ that can be deduced to be nonzero based on combinatorial analysis of each of its composing terms. The consequence of Corollary \ref{cor:sens_distr} is that if we take $\Sigma A^T\delta^{(k)}$ as an approximation to $\Sigma A^T(A\Sigma A^T)^{-1}\Delta$, we know which components of the approximation are guaranteed to be zero, i.e., are invariant to locally supported perturbations in the constraint vector.
Based on the previous theorem and  its corollary, we define a measure of distance between primal variables and dual variables that characterizes the indirect path, through coupling in the constraints, by which a perturbation in the constraint propagates to primal variables,
\begin{equation*}\label{eq:distance}
    d(v^{(p)}_i, v^{(d)}_j) := \min\{k \mid \mathcal{N}_1^{G_\text{opt}}(\mathcal{N}^{G_\text{con}}_{k - 1}(i))\}.
\end{equation*}
We also define the distance between sets of primal and dual variables as \[d(I,J) = \min\{ d(v^{(p)}_i, v^{(d)}_j)|{v^{(p)}_i\in I, v^{(d)}_j\in J}\}.\]
\paragraph{Component-wise sensitivity}
We will now show that the previous result along with the convergence guarantees of CR can be used to infer the component-wise magnitudes of the sensitivity expression. We will ultimately conclude that these magnitudes decay exponentially with rate $\frac{\sqrt{\kappa} - 1}{\sqrt{\kappa} + 1}$ with the degrees of separation between a component of $x$, and the support of $\Delta$, where $\kappa$ is the condition number of $A \Sigma A^T$.
\begin{theorem}[Decay in Sensitivity]\label{thm:decay_sens}
The component-wise magnitudes of the sensitivity expression can be bounded as 
\[\norm{[D(b)\Delta]_S}_2 \leq C \left(\frac{\sqrt{\kappa} - 1}{\sqrt{\kappa} + 1}\right)^{d(S, \text{supp}(\Delta))},\]
where $C = \frac{2\norm{\Delta}_2}{\sigma_\text{min}(A)}$, and $\kappa = \dfrac{\lambda_{\text{max}}(A \Sigma A^T)}{\lambda_{\text{min}}(A \Sigma A^T)}$ is the condition number of $A \Sigma A^T$.
\begin{proof}
Let $\delta^{(k)}$ be the $k$th estimate of $(A\Sigma A^T)^{-1}\Delta$ generated via the Conjugate Residuals algorithm. Corollary \ref{cor:sens_distr} allows us to conclude that $[\Sigma A^T \delta^{(k)}]_S = 0$ if $k \leq {d(S, \text{supp}(\Delta))}$.
It then follows that for all $k \leq d(S, \text{supp}(\Delta)) $
\begin{align*}
    [D(b)\Delta]_S &= [D(b)\Delta - \Sigma A^T \delta^{(k)}]_S\\
    &= [\Sigma A^T((A\Sigma A^T)^{-1}\Delta - \delta^{(k)})]_S.
\end{align*}
Taking the norm of both sides of the equality, we can bound the sensitivity as \[\norm{[D(b)\Delta]_S}_2 \leq \norm{\Sigma A^T((A\Sigma A^T)^{-1}\Delta - \delta^{(k)})}_2.\]
\frmargin{Notice that}{This section should be polished a bit once the content is stable.} the $k$th residual can be expressed as \[r_k =  A\left(\Sigma A^T \left((A\Sigma A^T)^{-1} \Delta - \delta^{(k)}\right)\right),\]
and convergence of the conjugate residuals algorithms guarantees that 
\[ \norm{r_k}_2 \leq 2\left(\frac{\sqrt{\kappa} - 1}{\sqrt{\kappa} + 1}\right)^k \norm{r_0}_2.\]
Consequently, using the fact that $\sigma_{\text{min}}(A) \norm{v}\leq\norm{Av}$, we can bound
% \frmargin{Therefore, we can bound}{$\sigma_\text{min}(\cdot)$ is not defined! Add to definitions and notation.}
\begin{align*}
    \norm{[D(b)\Delta]_S}_2 &\leq \norm{\Sigma A^T((A\Sigma A^T)^{-1}\Delta - \delta^{(k)})}_2\\
    &\leq \frac{\norm{r_k}_2}{\sigma_\text{min}(A)} \leq \frac{2\norm{\Delta}_2}{\sigma_\text{min}(A)}\left(\frac{\sqrt{\kappa} - 1}{\sqrt{\kappa} + 1}\right)^k.
\end{align*}
Taking $C = \frac{2\norm{\Delta}_2}{\sigma_\text{min}(A)}$ and $k = d(S, \text{supp}(\Delta))$ concludes the proof.\end{proof}
\end{theorem}
Theorem \ref{thm:decay_sens} states that components that are ``closest" to the perturbation, i.e., those that participate in the constraints, are most sensitive to the perturbation, and the sensitivity of components decay exponentially according to their degree of separation from the perturbation. Moreover, the decay rate can be bounded by $\frac{\sqrt{\kappa} - 1}{\sqrt{\kappa} + 1}$. Theorem \ref{thm:decay_sens} can be readily extended to bound the effect that perturbations in the constraint vector, $b$, have on individual components of the correction factor.
\begin{corollary}[Decay in Error]\label{cor:dist_corr}
If $\lambda \geq \frac{\sqrt{\kappa(x)} - 1}{\sqrt{\kappa(x)} + 1}$ for all $x$, then 
\[\norm{\left[x^*(\hat{b} + \Delta) - x^*(\hat{b})\right]_S} \leq C \lambda^{d(S, \text{supp}(\Delta))},\]
for $C = \frac{2\norm{\Delta}}{\sigma_\text{min}(A)}$.
\begin{proof}
Like before, we define $x_\theta := x^*(\hat{b} + \theta\Delta)$, and $b_\theta := \hat{b} + \theta\Delta$. Then, 
\begin{align*}
    &\norm{\left[x^*(\hat{b} + \Delta) - x^*(\hat{b})\right]_S}\\ &= \norm{\left[\int_0^1 \Sigma(x_\theta)A^T(A\Sigma(x_\theta)A^T)^{-1}\Delta d\theta\right]_S}\\
    &= \norm{\int_0^1[D(b)\Delta]_Sd\theta} \leq \int_0^1\norm{[D(b)\Delta]_S}d\theta\\
    &\leq \int_0^1 \norm{\Sigma(x_\theta) A^T((A\Sigma(x_\theta A^T)^{-1}\Delta - \delta^{(k)})} d\theta\\
    &\leq \int_0^1 \frac{2\norm{\Delta}}{\sigma_\text{min}(A)}\left(\frac{\sqrt{\kappa(x_\theta)} - 1}{\sqrt{\kappa(x_\theta)} + 1}\right)^k d\theta \leq \frac{2\norm{\Delta}}{\sigma_\text{min}(A)}\lambda^k.
\end{align*}
Taking $C = \frac{2\norm{\Delta}}{\sigma_\text{min}(A)}$ completes the proof. 
\end{proof}
\end{corollary}

% Like before, we define $x_\theta := x^*(\hat{b} + \theta\Delta)$, and $b_\theta := \hat{b} + \theta\Delta$. Then, 
% \begin{align*}
%     &\norm{\left[x^*(\hat{b} + \Delta) - x^*(\hat{b})\right]_S}\\ &= \norm{\left[\int_0^1 \Sigma(x_\theta)A^T(A\Sigma(x_\theta)A^T)^{-1}\Delta d\theta\right]_S}\\
%     &= \norm{\int_0^1[D(b)\Delta]_Sd\theta} \leq \int_0^1\norm{[D(b)\Delta]_S}d\theta\\
%     &\leq \int_0^1 \norm{\Sigma(x_\theta) A^T((A\Sigma(x_\theta A^T)^{-1}\Delta - \delta^{(k)})} d\theta\\
%     &\leq \int_0^1 \frac{2\norm{\Delta}}{\sigma_\text{min}(A)}\left(\frac{\sqrt{\kappa(x_\theta)} - 1}{\sqrt{\kappa(x_\theta)} + 1}\right)^k d\theta \leq \frac{2\norm{\Delta}}{\sigma_\text{min}(A)}\lambda^k.
%     % &\leq \int_0^1 2\sqrt{\kappa(x_\theta)}\norm{\Sigma(x_\theta) A^T}_2\norm{(A\Sigma(x_\theta) A^T)^{-1}\Delta}_2\left(\frac{\sqrt{\kappa(x_\theta)} - 1}{\sqrt{\kappa(x_\theta)} + 1}\right)^k d \theta\\
%     % &\leq \lambda^k \int_0^1 2\sqrt{\kappa(x_\theta)}\norm{\Sigma(x_\theta) A^T}_2\norm{(A\Sigma(x_\theta) A^T)^{-1}\Delta}_2 d\theta
%     % &= \norm{\int_0^1 \mathds{1}_S\Sigma(x_\theta)A^T(A\Sigma(x_\theta)A^T)^{-1}\Delta d\theta}\\
%     % &\leq \int_0^1 \norm{\mathds{1}_S\Sigma(x_\theta)A^T(A\Sigma(x_\theta)A^T)^{-1}\Delta} d\theta\\
% \end{align*}
% Taking $C = \frac{2\norm{\Delta}}{\sigma_\text{min}(A)}$ completes the proof. 

Corollary \ref{cor:dist_corr} extends the results of Theorem \ref{thm:decay_sens} to establish that the magnitude of the correction factor decays with distance to the perturbation. The authors of \cite{8370673} characterized a similar decay bound for network flow problems, and demonstrated the potential of such a bound in the context of warm-start optimization. This decay bound extends their results to all linearly-constrained convex optimization problems, and improves on our previous results derived from the infinite series expansion of the sensitivity expression \cite{BrownRossiea20}.

\subsection{Putting it all together}\label{subsec:summary}
We now have the technical machinery necessary to establish a notion of locality. In this section, we restrict our attention to local sub-problems induced by a $k$-hop neighborhood around $x_i$ in $G_\text{dec}$. To lighten notation, we let $x^{(k)}$ denote the solution to the local sub-problem induced by the $k$-hop neighborhood around $i$ (denoted by $x^{\left(\mathcal{N}^{G_\text{dec}}_k(i)\right)}_i$ in Section \ref{subsec:local_global}).
In this section, we show that \[\lvert x^{(k)}_i - x^*_i \rvert \leq C \lambda^k,\]
for constants $C$ and $\lambda$, and provide bounds on both $C$ and $\lambda$. In other words, we will show that the error in component $i$ decays exponentially according to rate $\lambda$ with the size of neighborhood generating the local sub-problem. The rate $\lambda$ naturally characterizes the degree to which local information is sufficient to compute a single component of the global optimum, ultimately, becoming our metric of ``locality".

We proceed by leveraging the results of Section \ref{subsec:local_global} to characterize the error on each of the local sub-problems in terms of the implicit constraints, $\hat{b}^{(k)}$. We will then apply the results derived in Section \ref{subsec:sensitivity} to bound the error induced at component $x_i$. The key difficulty resolved in this section stems from the fact that we want to avoid solving for the implicit constraints (which would require using the entirety of the problem, thus defeating the purpose of locality!)---this is akin to applying Corollary \ref{cor:dist_corr} without knowing $\Delta$. 

While we generally cannot control the value of the implicit constraints, $\hat{b}^{(k)}$, the construction of the local sub-problems guarantees that the distance from $i$ to the cut constraints is at least $k$, i.e., $d(i, \text{supp}(\Delta^{(k)})) \geq k $ where $\Delta^{(k)} := b - \hat{b}^{(k)}$. Moreover, we know that the ``perturbations", $\Delta^{(k)}$, are not arbitrary---they arise from ``cutting" constraints. These insights provide sufficient knowledge of $\Delta^{(k)}$ to apply Corollary \ref{cor:dist_corr}. We are now in a position to prove the main result.% of this paper.
\begin{customthm}{\ref{thm:locality_sketch}}
% Let $x^{(k)}_i$ denote the component $i$ of the solution to the local sub-problem induced by a $k$-hop neighborhood around $x_i$. Then
% \[\lvert x^{(k)}_i - x^*_i \rvert \leq C \lambda^k\]
% for some constant $C$, if $\lambda \geq \frac{\sqrt{\kappa(x)} - 1}{\sqrt{\kappa(x)} + 1}$ for all $x$.
Let $x^{(k)}$ be the solution to the optimization problem induced by restricting Problem \ref{global_opt} to $k$-hop neighborhood around $x_i$,  $\mathcal{N}_k^{\text{(dec)}}(i)$, and the constraints only involving those variables. If $\lambda = \sup_x \frac{\sqrt{\kappa(x)} - 1}{\sqrt{\kappa(x)} + 1}$, where $\kappa(x)$ denotes the condition number of $A \nabla^2 f(x)^{-1} A^T$, then
\begin{equation}
\lvert x^{(k)}_i - x^*_i \rvert \leq C \lambda^k
\label{eq:bound_x}
\end{equation}
for $C = \left(1+ \sqrt{\frac{L}{\mu}}\right)\frac{2\sigma_{max}(A)}{\sigma^2_{min}(A)}\norm{ b - Ax^*_{UC}}_2$.
\begin{proof}
First, we will show that the $k$-hop local sub-problem can be generated by cutting constrains that are at least distance $k$ from $i$ under the primal-dual distance metric. We will prove this by reasoning about the supports of the appropriate matrix products. The set of primal variables contained in the $k$-hop neighborhood of $x_i$ can be equivalently characterized as
\[\mathcal{N}^{(p)}_k(i) = \left\{j \mid \left[(A^TA)^k\right]_{ij} \neq 0\right\} = \text{supp}([(A^TA)^k]_{i*}).\]
Similarly, the primal-dual distance metric can be defined as 
\begin{align*}
d(i, c) &=\min\{k \mid c \in \text{supp}\left(\left[A^T(AA^T)^{k - 1}\right]_{i*}\right)\}\\
&=\min\{k \mid c \in \text{supp}\left(\left[(A^TA)^{k - 1}A^T\right]_{i*}\right)\}.
\end{align*}
Because the graph $G_\text{dec}$ is defined by placing an edge between agents that appear together in the same constraint, if $A_{c, i} \neq 0$ and $A_{c, i} \neq 0$ for some constraint $c$, then for all $l \in V^{(d)}$, \[\vert d(i, l) - d(j, l) \vert \leq 1.\]
Moreover, to generate the $k$-hop local sub-problem, a constraint only cut if it contains a variable of distance at least $k + 1$. Consequently, all of the primal variables in the cut constraint are at least distance $k$ from $i$.
We can now apply Corollary \ref{cor:dist_corr} to bound the error in component $i$ as \[\lvert x^{(k)}_i - x^*_i\rvert \leq \frac{2\norm{\Delta^{(k)}}}{\sigma_\text{min}(A)} \lambda^k.\]

We will bound the $\Delta^{(k)}$ term by deriving the maximum constraint violation error, $\norm{b - \hat{b}^{(k)}}_\infty$. We do so by noting that the solution to the local sub-problems are consistent with the solution to 
\begin{argmini}|l|
{x\in \mathbb{R}^N}{f(x)}
{}{\hat{x}^{\mathcal{N}(i, k)} = }
\addConstraint{A_{C_{\mathcal{N}(i, K)}}x = b_{C_{\mathcal{N}(i, K)}}}.
\end{argmini}
That is, we use the same set of constraints as agent $i$'s $k$-hop local sub-problem but include all of the variables in the objective function. Precisely, 
\[x^{(\mathcal{N}(i, k))} = \left[\hat{x}^{\mathcal{N}(i, k)}\right]_{\mathcal{N}(i, K)}.\]
Consequently, every variable but those in $\mathcal{N}(i, k)$ is unconstrained. 
We define
\begin{argmini}|l|
{x\in \mathbb{R}^N}{f(x)}
{}{x^*_{UC} = }
\end{argmini}
to be the solution to the unconstrained problem. Then
\begin{equation*}
\left[\hat{x}^{\mathcal{N}(i, K)}\right]_i = 
\begin{cases}
x^{(\mathcal{N}(i, k))}_i, &\text{if } i  \in \mathcal{N}(i, k)\\
\left[x^*_{UC}\right]_i, &\text{if } i \not \in \mathcal{N}(i, k).
\end{cases}   
\end{equation*}
Then, the individual components of the implicit constraints can be derived as 
\begin{equation*}
\left[\hat{b}^{(k)}\right]_i = 
\begin{cases}
b_i, &\text{if } i \in C_{\mathcal{N}(i, k)}\\
\left[Ax^{(\mathcal{N}(i, k))}\right]_i, &\text{if } i \not \in C_{\mathcal{N}(i, k)}.
\end{cases}   
\end{equation*}
It then follows that the component-wise constraint violation is given by
\begin{equation*}
\left[b - \hat{b}^{(k)}\right]_i = 
\begin{cases}
0, &\text{if } i \in C_{\mathcal{N}(i, K)}\\
\left[b - A\hat{x}^{(\mathcal{N}(i, K))}\right]_i, &\text{if } i \not\in C_{\mathcal{N}(i, K)}\\
\end{cases}   
\end{equation*}
Consequently, the maximum constraint violation is equal to 
\[\norm{[b - A\hat{x}^{(\mathcal{N}(i, K))}}_\infty.\]
To obtain a uniform bound, we will show that \[\norm{b - A\hat{x}^{(\mathcal{N}(i, K))}}_2\leq 
\left(1+ \sqrt{\frac{L}{\mu}}\right)\frac{\sigma_{max}(A)}{\sigma_{min}(A)}\norm{ b - Ax^*_{UC}}_2\]
Because $f$ is $L$-smooth and $\mu$-strongly convex,
\begin{align*}
    \frac{\mu}{2}\norm{\hat{x} - x_{UC}}^2_2 &\leq f(\hat{x}) - f(x_{UC}) &&\leq \frac{L}{2}\norm{\hat{x} - x_{UC}}^2_2\\
     \frac{\mu}{2}\norm{x - x_{UC}}^2_2 &\leq f(x) - f(x_{UC}) &&\leq \frac{L}{2}\norm{x - x_{UC}}^2_2
\end{align*}
Because $f(\hat{x}) \leq f(x)$, \[\frac{\mu}{2}\norm{\hat{x} - x_{UC}}^2_2 \leq \frac{L}{2}\norm{x - x_{UC}}^2_2.\]
Then, using the triangle inequality, 
\begin{align*}
    \norm{x - \hat{x}}_2  &= \norm{x - x_{UC} + x_{UC} - \hat{x}}_2\\
    &\leq \norm{x - x_{UC}}_2 + \norm{x_{UC} - \hat{x}}_2\\
    &\leq \left(1+ \sqrt{\frac{L}{\mu}}\right)\norm{x - x_{UC}}_2
\end{align*}
Finally, because $\sigma_{min}(A)\norm{v} \leq \norm{Av} \leq \sigma_{max}(A)\norm{v}$ and $b = Ax$,\[\norm{b - A\hat{x}}_2 \leq \left(1+ \sqrt{\frac{L}{\mu}}\right)\frac{\sigma_{max}(A)}{\sigma_{min}(A)}\norm{ b - Ax^*_{UC}}_2.\]
\end{proof}
\end{customthm}
The upshot of this theorem is that if an accuracy of ${\lvert x^{(k)}_i - x^*_i \rvert \leq \epsilon}$ is desired, a neighborhood size of  \[K \geq \frac{1}{1-\lambda}\log\left(\frac{C}{\epsilon}\right)\] \frmargin{is sufficient}{High-level comment. At this point, the reader may be wondering "How big is C? How useful is this bound?" This is a great place to point the reader to say "In Section \ref{sec:experiments:economic:convergence}, we show through numerical simulations that the proposed bound is remarkably tight for highly local problems.}. 
The larger $\lambda$ is, the larger the neighborhood needed to achieve a desired accuracy, whereas a smaller $\lambda$ indicates that a smaller neighborhood is sufficient. We note here that the actual number of variables and constraints included in a neighborhood of a fixed size will depend on the problem. For example, if $G_\text{dec}$ is a path graph, then the number of variables  in each neighborhood will scale linearly with $k$, whereas if $G_\text{dec}$ is a grid graph, then the number of variables in each neighborhood scales quadratically with $k$. 

The close relationship between $\lambda$ and the size of sub-problem needed to achieve a desired accuracy justifies it as a metric of the degree to which local information is sufficient to approximate individual components of the global solution. We are now in a position to define our metric of locality. 
\begin{definition}[Locality]\label{def:locality}
For an optimization problem of the form \eqref{global_opt} we define the locality of the problem as \begin{equation}\lambda(f, A) = \sup_x \frac{\sqrt{\kappa(x)} - 1}{\sqrt{\kappa(x) + 1}}.\end{equation}
We also extend the definition of locality to classes of problems. Explicitly, if it is known that $f \in F$ and $A \in \mathcal{A}$, we define the locality of the class of problems as
\begin{equation}
\lambda(F,\, \mathcal{A}) = \sup_{f \in F, \, A \in \mathcal{A}} \lambda(f,\, A). 
\end{equation}
\end{definition}
For instance, in network flow problems the class of constraint matrices, $\mathcal{A}$, are those representing flow conservation constraints. The flow conservation constraint at a given node only affects variables for flows departing or arriving at that node; accordingly, the distance metric $d$ corresponds to the shortest-path distance in the network flow graph. 

\subsection{Discussion}
% Outline:
% \begin{enumerate}
%     \item Static case: In scenarios where the objective function $f$ and the constraint matrix remain fixed, the locality parameter can be computed up front and passed as a parameter to the network. For example, 
    
%     \item Classes cases: To generalize to problem instances that might exhibit variation in the objective and constraints matrix, we can still reason about the locality in terms of def. For example,... In the 
%     \item Complementary notion of stochastic metric of locality
%     \item Motivates adaptive schemes 
% \end{enumerate}
In this section, we have proposed a metric of locality that captures the amount of information that is required to solve for a single component of a convex optimization problem to a given degree of accuracy. From a practical standpoint, implementing the locality-aware algorithm requires checking the condition number for a given problem instance. In scenarios where the objective function, $f$, and constraint matrix, $A$, are fixed (with potentially varying constraint vector $b$), the locality parameter can be computed once, offline, and passed in as a parameter to the network. As an example of such a setting, in Section \ref{sec:experiments} we consider an example of economic dispatch, in which we minimize an objective function capturing generation and transmission costs subject to load fulfillment constraints. In such a scenario, the objective function and constraint matrix are fixed while the constraint vector is determined online. Since the objective function and constraint matrix are static, the proposed results can be immediately applied.

In Definition \ref{def:locality}, we generalize our metric of locality to classes of problems to account problem instances that exhibit variability in the objective and constraint matrix. As an example, in Section \ref{sec:experiments} we consider an instance of the power network state estimation problem, in which we maximize the posterior probability of the power flows and voltage angles given noisy measurements of both, subject to the power flow equations. The class of problems encompassing this scenario is defined by objective functions derived from the maximum-a-posteriori estimation formulation, and the constraint matrix representing the power flow equations. The noisy measurements are modeled in the objective function, so, in contrast with the economic dispatch example, the objective function is stochastic and determined at run-time. We show that the Hessian of the objective function is constant for all possible objective functions of this form. Accordingly, the locality metric can be readily computed in this setting. However, we remark that this is not always be the case, and there is often a practical trade-off between generality of a class of problems and how informative our metric of locality is. For example, if all but one problem in a class exhibit a high degree of locality, the proposed metric would still indicate that the entire class exhibits a low degree of locality---resulting in bounds that are exceedingly conservative for almost all of the problems in that class.

In the case that computing the locality of an entire class of problem is intractable, we suggest a sampling-based approach, where individual problem instances are sampled, and their locality estimated. This motivates a complementary notion of locality in a stochastic sense, where the presented notion of locality is extended from being a worst-case bound to one that captures the distribution of locality parameters in a class of problem. Similarly, we highlight the potential for a class of adaptive algorithms where agents individually estimate local measures of locality based on problem data within their neighborhood (potentially by applying notions of structured and component-wise condition numbers \cite{gohberg1993mixed}). This not only would alleviate the overhead of computing the global locality parameter, but would remedy the inherent conservatism of worst-case bounds---as demonstrated in Section \ref{sec:experiments}, the maximum error of Algorithm \ref{alg:locality} across agents can be much worse than the average error.

\section{Numerical Experiments}\label{sec:experiments}
In this section we empirically validate our theoretical bounds and assess the performance of the locality-aware algorithm. 

First, we consider a synthetic instance of the economic dispatch problem. We compare the true error of the locality-aware algorithm with the theoretical upper-bound on the error, as a function of the number of communication rounds. We observe that when the condition number is relatively low,  the performance of the algorithm closely matches the theoretical prediction.  We also assess the performance of the projected sub-gradient algorithm and observe that the number of iterations necessary to achieve a high level of accuracy far exceeds the number of communication rounds required for the  locally-aware algorithm. 

Second, we consider the state-estimation problem on the Pan European Grid Advanced Simulation and State Estimation (PEGASE) 9241-bus power-network \cite{6488772}, \cite{josz2016ac}. From a theoretical standpoint, this problem exhibits a high locality rate, which suggests that a locality-aware algorithm will not be useful in this case. However, empirically we observe that the locally-aware algorithm still manages to find a high-quality solution in fairly few rounds. This suggests that the locality rate is overly conservative for this case.

Finally, we consider a randomized instance of the rendezvous problem. Intuitively, deciding on a meeting location that is central to all agents is an inherently global problem. This is confirmed by the high locality parameter. In contrast to the state-estimation example, the rendezvous problem does not exhibit locality that is overlooked by the theory. This confirms that our characterization of locality does not buy us locality when there is none.
\subsection{Economic Dispatch}
\subsubsection{Problem Setting}
We consider a setting where generators are positioned in an $N \times M$ grid, and load buses are positioned in the center of each grid cell. Each load bus is only connected to its neighboring generators, which need to supply enough power to satisfy a stochastically generated load $\mathcal{L}(i)$. The costs associated with the problem are a quadratic generation cost with coefficient $\frac{\alpha}{2}$, and a quadratic transmission cost with coefficient $\frac{\beta}{2}$.
Explicitly, the optimization problem representing this setting is given by
\begin{mini}|l|
{x}{\frac{\alpha}{2}\sum_i \left(\sum_{j \in \mathcal{N}(i)} x_{i,j}\right)^2 + \frac{\beta}{2}\sum_i  \sum_{j \in \mathcal{N}(i)} x_{i,j}^2}
{\label{ex:econ_disp}}{}
\addConstraint{\sum_{i \in \mathcal{N}(j)} x_{i,j}= \mathcal{L}_j, \, \forall j.}
\end{mini}
If $\alpha = 0$, the problem becomes fully decoupled and the optimal solution is given by splitting each load evenly between its generators. Consequently, this setting allows us to use the parameters $\alpha$ and $\beta$ to ``tune'' the locality of the problem and investigate both the tightness of the proposed bounds, and the performance of the locality-aware algorithm for various rates of locality. We note that this example also illustrates the extension of our results to \frmargin{block-separable}{Did you define this anywhere? It is quite clear from context, but I lean towards pedantry} objectives. 
\subsubsection{Effect of Locality on Convergence}\label{sec:experiments:economic:convergence}
\begin{figure*}[h!]
    \centering
    \includegraphics[width = 0.8\textwidth]{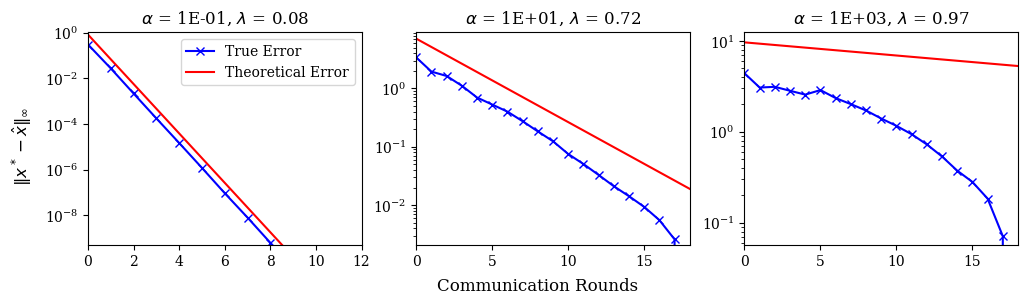}
    \caption{This figure plots the true accuracy of the locality-aware algorithm (in blue) against the theoretical accuracy (in red) for varying communication rounds. In the well-conditioned case, $\alpha = 0.1$, the proposed theoretical rate is tight. As the conditioning of the problem increases, the theoretical bound becomes more conservative. }
    \label{fig:econ_disp_locality}
\end{figure*}

In this example, we fixed the dimension of the global problem to be $20 \times 20$, and varied $\alpha$ to be $0.1$, $10$, and $1000$. The condition number for each of these cases was calculated and found to be $1.39$, $37.62$, and $3611.43$ respectively---these correspond to locality parameters of 0.08, 0.72, and 0.97. In each of these cases, we varied the local sub-problem size for each of the agents between $0$ and the diameter of the network. Figure \ref{fig:econ_disp_locality} plots the maximum error (computed over all the agents) against the size of local sub-problem, as well as the error bound in Theorem \ref{thm:locality_sketch} derived from the locality parameter. For well-conditioned problems, the true performance of the algorithm aligns closely with the theoretical prediction, while the theoretical bounds become more conservative as the condition number and the locality parameter increase. Notably, in cases with low locality parameter, the error exhibits clear exponential convergence. Whereas, when the locality parameter is higher, the convergence rate of the error appears to increase with the number of communication rounds. This aligns closely with the superlinear convergence behavior observed when solving large symmetric systems of equations using Krylov subspace methods \cite{Beckermann.ea.SIAM01}.

\subsubsection{Comparison to other methods}
We now evaluate the performance of our algorithm against the standard distributed projected subgradient algorithm \cite{5404774}. The distributed subgradient algorithm assumes an optimization problem of the form
\begin{mini}|l|
{x\in \mathbb{R}^N}{\sum_{i = 1}^m f_i(x)}
{}{}
\label{eq:subgrad_form}
\addConstraint{x \in \chi_i}
\end{mini}
where each $f_i(x)$ and $\chi_i$ are only known by agent $i$, and messages are passed over a fixed communication topology.

\begin{figure*}[h!]
    \centering
    \includegraphics[width = 0.8\textwidth]{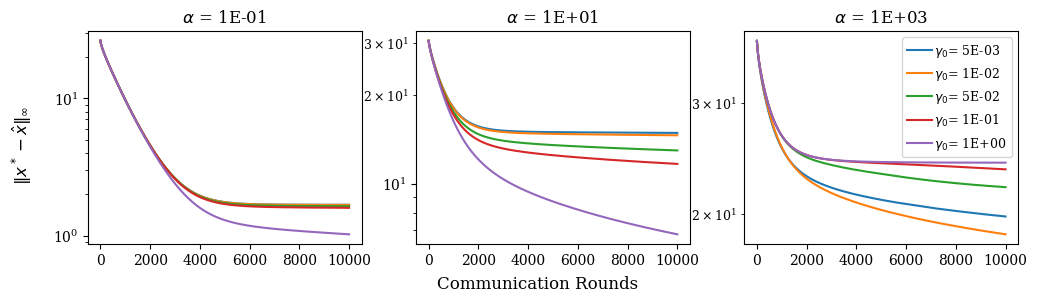}
    \caption{This figure plots the convergence of the projected sub-gradient algorithm, with lazy Metropolis weighting, against the number of communication rounds for varying intial step-sizes. The convergence of the algorithm is highly sensitive to the initial step-size.}
    \label{fig:econ_disp_projSG}
\end{figure*}

Each generator's local objective function encodes its own transmission and generation costs, i.e., \begin{equation*}f_i(x) = \frac{\alpha}{2}\left(\sum_{j \in \mathcal{N}(i)} x_{i,j}\right)^2 + \frac{\beta}{2}\sum_{j \in \mathcal{N}(i)} x_{i,j}^2,\end{equation*}
and each generators' local constraint sets are the load constraints it needs to satisfy. We assume a fixed communication graph where each generator can communicate with other generators that it shares a constraint with---this is exactly the communication graph assumed in Algorithm \ref{alg:locality}. 
We use the lazy Metropolis weighting for the consensus step (let $L$ denote the matrix encoding these weights). Every agent maintains and updates a copy of the global variable during each iteration. Let $x^k_{(i)}$ denote the $i$th agent's copy of the global optimization variable at iteration $k$. Then the projected subgradient updates are given by 
\[x^{k + 1}_{(i)} = \Pi_{\chi_i}\left(\sum_{j} L_{ij}x^{k}_{(j)} - \frac{\gamma_0}{k^{0.55}}g^k_{(i)}\right),\]
where $\Pi_{\chi_i}(x)$ is the orthogonal projection of the point $x$ on the set $\chi_i$.
%given by, 
%\[x^{k+1}_i = x^k_i + \sum_{j = \mathcal{N}^k_i} \frac{1}{2\max\{d^k_i,d^k_j\}}(x^k_j - x^k_i)\]
We simulated the projected subgradient algorithm for varying values of $\alpha$ for 10,000 iterations, and extracted local estimates from each agents' copy of the global decision variable i.e., $\hat{x}^k_i = \left[x^{k}_{(i)}\right]_i$. Figure \ref{fig:econ_disp_projSG} plots the maximum error across all agents of the projected sub-gradient algorithm against the number of communication rounds. We observe that within 10,000 iterations, none of the estimates have converged to the error achieved by the initial communication round in the locality-aware algorithm despite each agent having access to all of the problem data it would have obtained after the initialization round. %\rbline{}{This suggests that a hybrid between the locality-aware approach, i.e., one where we initialize with some number of communication rounds of the locality algorithm, and then use project sg to fine tune after may drastically speed up projSG while combatting the growing message size problem.}

We also note that the convergence of the projected subgradient algorithm is particularly sensitive to the step-size schedule, and that the optimal step size is dependent on the condition number of the problem. Moreover, the best initial step size is not consistent across different problem instances---for $\alpha = 10$, an initial steps-size of $\gamma_0 = 1$ converged the fastest, whereas for $\alpha = 1000$, an initial step-size of $\gamma_0 = 0.01$ converged the slowest. While it is a weakness that the locality-aware algorithm depends on the locality parameter, which depends on the condition number, efficient implementation of the projected sub-gradient algorithm depends on the condition number as well.

\subsection{Power Network---State Estimation}
\begin{figure}
    \centering
    \includegraphics[width=0.48\textwidth, height = 0.3\textwidth]{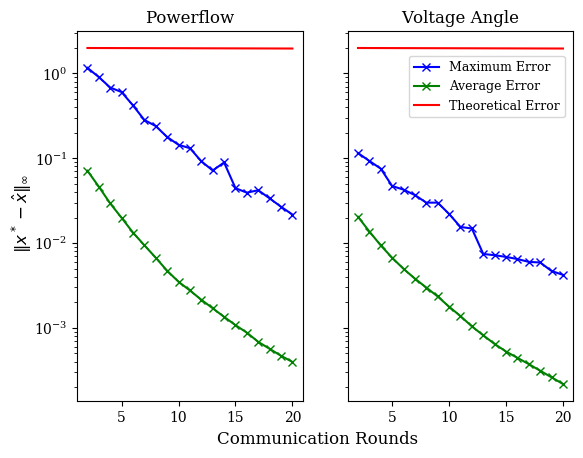}
    \caption{This figure depicts the local sub-problem size versus average (green), maximum (blue), and theoretical (red) errors in power flow and voltage angle estimates. The theoretical bounds suggest a rate of decay of 0.9992. However, both the maximum and average errors decay much faster, with the average error being consistently an order of magnitude smaller than the theoretical error.}
    \label{fig:pegase9241}
\end{figure}

We consider a power network modeled by a graph $G(V, E)$. We assume that the network is primarily inductive, the voltage amplitudes are fixed to one, and the voltage angle differences between neighboring nodes are small enough to apply the DC power assumption. The power flow $P_{ij}$ on edge $\{i,\, j\} \in E$ must satisfy the equation \[P_{ij} = -b_{ij}(\theta_i - \theta_j).\]
We consider a setting where both the voltage angles, $\theta$, and line power flows, $P$,  are measured according to \[\theta^m_i = \theta_i + \epsilon_i, \quad P^m_{ij} = P_{ij} + \epsilon_{ij}\]
where $\epsilon_i \sim \mathcal{N}(0, \sigma^2_i)$, and $\epsilon_{ij} \sim \mathcal{N}(0, \sigma^2_{ij})$, and the true power flow and voltage angles must be estimated. Then, the maximum a posteriori estimation problem is given by 
\begin{mini}|l|
{\hat{\theta}\in \mathbb{R}^{|V|}, \,  \hat{P}\in \mathbb{R}^{|E|}}{\sum_{i \in V} \left(\frac{\hat{\theta}_i -\theta^m_i}{\sigma_i}\right)^2 + \sum_{(i,j) \in E}\left(\frac{\hat{P}_{ij} - P^m_{ij}}{\sigma_{ij}}\right)^2}
{\label{pn_opt}}{}
\addConstraint{\left[
\begin{array}{c|c}I & B \end{array}
\right] \left[\begin{array}{c}
  \hat{P}\\\hline\hat{\theta}
\end{array}\right]}{= 0}
\end{mini}
where $I$ is the identity matrix, and $B$ is the network admittance matrix containing the electrical parameters and topology information \cite{Molzahn.ea.17}. We simulated the locality-aware distributed optimization algorithm (Algorithm \ref{alg:locality}) for $K = 2, \ldots, 20$. The average and maximum errors in both the powerflow and voltage angle estimates are shown in Figure \ref{fig:pegase9241} along with their theoretical bounds. We found that the condition number of the problem was $6.37 \times 10^6$, resulting in a locality rate of $0.9992$. The theoretical bounds, in this case, would suggest that the locality-aware approach is not well-suited to the problem setting. However, numerically, we observe that this bound is overly conservative and the problem instance nevertheless exhibits locality behavior. Additionally, we see that the average error tends to be an order of magnitude less than the maximum error exhibited. Our method of analysis resulted in a uniform worst-case bound, however, this experiment demonstrates that the worst case is a poor representation of the average case. Accordingly, we highlight extending the results of this paper to quantify local measures of locality.

% \frmargin{local measures of locality could potentially alleviate conservatism inherent to any global bound.}{This sentence is suggestive but esoteric. Please add ~1 sentence on how you would do this at a high level. (I will steal it for the ONR proposal).}

\subsection{Rendezvous}
We now consider an instance of rendezvous where 1000 agents are places randomly in a $[0,1] \times [0, 1]$ grid, and they must decide on a meeting location the minimizes the sum of their distances to the location. The optimization problem representing this setting is given by
\begin{mini}|l|
{x, y \in \mathbb{R}}{\sum_{i = 1}^N (x - x_i)^2 + (y - y_i)^2}
{}{}
\label{eq:uncon_rendezvou}
\end{mini}

\begin{figure}
    \centering
    \includegraphics[width=0.48\textwidth, height=0.3\textwidth]{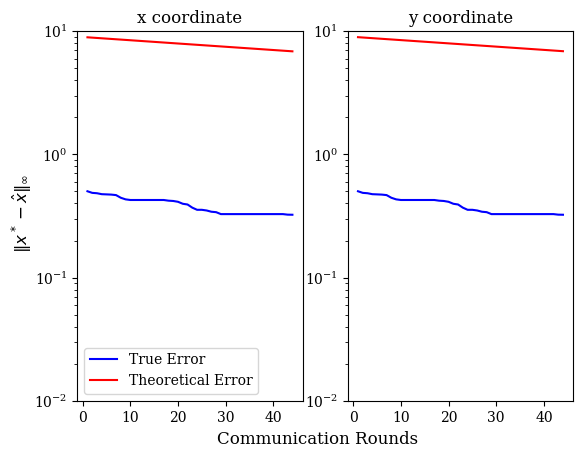}
    \caption{This figure shows the true accuracy of the locality-aware algorithm (blue) against its theoretical accuracy (red). The locality parameter, $\lambda = 0.9939$, indicates that the error should hardly decay with the number of communication rounds, which aligns with the empirical results observed.}
    \label{fig:rendezvous}
\end{figure}

We assume that the communication graph, $G = (V, E)$ between agents is a given by the minimum weight spanning tree of their distances. We rewrite the rendezvous optimization problem in the following form to make it amenable to distributed optimization algorithms,
\begin{mini}|l|
{\hat{x}, \hat{y} \in \mathbb{R}^N}{\sum_{i = 1}^N (\hat{x}_i - x_i)^2 + (\hat{y}_i - y_i)^2}
{}{}
\label{eq:con_rendezvou}
\addConstraint{\hat{x}_i = \hat{x}_j,\, \hat{y}_i = \hat{y}_j  \quad \forall (i, j) \in E}
\end{mini}
This formulation simply local copies of the meeting location coordinates, $x$ and $y$, and ensures that the neighbors agree on the same meeting location. Because the communication graph is connected, this condition ensures that all agents agree on the same meeting location. As we might expect, deciding on a meeting location that is central to \emph{all} agents is an inherently global problem. This is confirmed by the locality parameter, which was found to be $\lambda = 0.9939$. The true error along with our theoretical bounds are plotted in Figure \ref{fig:rendezvous}: unlike the example of state-estimation in a power network, the rendezvous example did not exhibit locality that was overlooked by the theory. We note that the communication graph in this example had a radius of 39---empirically, we see that the maximum error hardly changes even when multiple agents have already accumulated the entirety of the problem data. 

This experiment shows that our characterization of locality does not buy us locality when there is none. Some problems that we might solve with a multi-agent system are inherently global, requiring information from all of the nodes to solve the problem with reasonable accuracy---others are inherently local. The purpose of this paper is not to imbue all problems with locality, but rather to develop a metric that can distinguish between the two.

\section{Conclusion}\label{sec:conclusion}

% \rbline{ }{Generalizes to inequality constraints through the notion of active sets}
% \rbline{ }{Maybe discuss why we chose to focus on communication rounds vs computation. tbh we completely abstracted away the computational complexity. 1. because problem (1) is about as easy of an optimization problem to solve and 2. in distributed computing for commodity clusters, communication DOMINATES computation, and that's even when every commodity machine is hooked up to the router. Empirical data on this would be nice to put numbers to why we chose this problem }
% \rbline{ }{Rewrite following paragraph to reflect new structure of the paper}
% In this paper, we have studied the locality of linearly-constrained convex optimization problems
In this paper, we have studied the structure of linearly-constrained strongly-convex optimization problems, and proved that \emph{all} such problems exhibit locality. Our results hinge on the Conjugate Residuals algorithm, which allow us to relate the locality of a problem to its conditioning. The rate of locality derived from CR is $\frac{\sqrt{\kappa} - 1}{\sqrt{\kappa} + 1}$ is a significant improvement to the $\frac{\kappa - 1}{\kappa + 1}$ rate derived in previous work via the infinite Neumann expansion. We applied this notion of locality to design a distributed optimization algorithm that explicitly takes advantage of this fact, and demonstrated our algorithm in the context of both economic dispatch and state-estimation in a power network.

% and provided a method of tracking the cascading effects of a perturbation of the remainder of the network. This gave rise to a metric of locality suggesting that certain global optimization problem with ``local" structure can be solved on much smaller scales. We applied this notion of locality to design a distributed optimization algorithm that explicitly takes advantage of this fact, and demonstrated our algorithm in the context of both economic dispatch and state-estimation in a power network.

While the framework of locality seems like a promising direction for designing multi-agent systems that scale well with the number of agents, a number of key questions remain open. The first is the issue of determining the locality parameter of a problem---as stated, it is determined by a uniform bound on condition number that may be difficult to solve for. While we have provided a bound via the condition number of the sensitivity expression, the numerical experiments show that this bound can be conservative, especially in settings where the condition number is poor.
It is also of interest to determine the locality of a problem in a distributed fashion. The next question is how we can exploit locality without explicitly aggregating any problem data. One commonly cited reason for the necessity of distributed optimization algorithms is to avoid sharing information about privately known objectives and constraints---it remains open how to incorporate privacy in our approach.

\section*{Acknowledgments}
Part of this research was carried out at the Jet Propulsion Laboratory, California Institute of Technology, under a contract with the National Aeronautics and Space Administration.

\renewcommand{\baselinestretch}{0.88}
\bibliographystyle{IEEEtran}
{\small
\bibliography{locality.bib}
}

\ifextendedv
\section{Appendix}

\fi

\end{document}